\documentclass[aps,prl,numerical,superscriptaddress,showpacs,floatfix,reprint]{revtex4-1}
\usepackage{amsmath,amssymb,mathrsfs,bm}
\usepackage{url}
\usepackage{hyperref}
\hypersetup{
colorlinks=true,
linkcolor=blue,
anchorcolor =red,
citecolor=blue,
filecolor = red,
urlcolor=blue,
pdfauthor=author}
\usepackage{txfonts}
\usepackage[utf8]{inputenc}
\usepackage{amsmath}
\usepackage{graphicx}
\def\av<#1>{\left\langle\,#1\,\right\rangle}
\def\ev<#1>{\left\langle\,#1\,\right\rangle_{\rm{ev}}}

\begin{document}


\title{Probing the QCD Critical End Point with Finite-Size Scaling of Net-Baryon Cumulant Ratios}

\author{ Roy~A.~Lacey}
\email[E-mail: ]{Roy.Lacey@Stonybrook.edu}
\affiliation{Department of Chemistry, 
Stony Brook University, \\
Stony Brook, NY, 11794-3400, USA}

\date{\today}

\begin{abstract}
Finite-size scaling (FSS) is applied to net-baryon cumulant ratios $C_2/C_1$, $C_3/C_2$, $C_4/C_2$, $C_3/C_1$, and $C_4/C_1$ measured in Au+Au collisions over the Beam Energy Scan Phase~I range $\sqrt{s_{NN}}=7.7$--$200$~GeV to constrain the location and universality class of the QCD critical end point (CEP). Although finite-size and finite-time effects suppress non-monotonic signatures in unscaled data, the FSS analysis reveals a collapse of measurements from different beam energies and centralities onto universal scaling functions. All cumulant ratios collapse under a single, common set of critical exponents and exhibit divergence patterns characteristic o 3D Ising critical behavior. The scaling results indicate a CEP at $\sqrt{s}_{\rm CEP}\approx33.0$~GeV, corresponding to $\mu_{B,\rm CEP}\approx130$~MeV and $T_{\rm CEP}\approx158.5$~MeV. These findings demonstrate that finite-size scaling provides a robust, model-independent framework for accessing critical behavior in finite, dynamically evolving systems, where non-equilibrium baryon-number transport can enhance the experimental visibility of susceptibility-driven fluctuations without modifying the underlying universality class.
\end{abstract}


\pacs{25.75.-q, 25.75.Dw, 25.75.Ld} 
\maketitle

%


The search for the Quantum Chromodynamics (QCD) critical end point (CEP) is a 
central objective in heavy-ion physics, as it offers a unique window into the 
phase structure of strongly interacting matter under extreme conditions. The 
CEP is expected to mark the termination of a first-order phase transition line 
and the onset of a smooth crossover in the QCD phase diagram, analogous to the 
liquid--gas critical point in conventional fluids~\cite{Stephanov:2004wx, 
Rajagopal:2000wf}. In equilibrium thermodynamic systems, proximity to such a 
critical point is characterized by universal scaling behavior governed by the 
growth of a correlation length and the associated critical exponents.

Experimentally, the CEP is probed in relativistic heavy-ion collisions using a 
variety of observables that are sensitive, in different ways, to changes in the 
underlying phase structure of QCD matter. Among these, event-by-event 
fluctuations of conserved charges—most notably the net-baryon number—play a 
central role and are a primary focus of the Beam Energy Scan (BES) program at 
the Relativistic Heavy Ion Collider (RHIC)~\cite{STAR:2020tga}. These fluctuations 
are quantified using cumulants of the net-baryon multiplicity distribution, 
including the mean ($C_1$), variance ($C_2$), skewness ($C_3$), and kurtosis 
($C_4$), as well as ratios such as $C_2/C_1$, $C_3/C_2$, $C_4/C_2$, $C_3/C_1$, and 
$C_4/C_1$. In the absence of critical phenomena, these ratios are expected to 
vary smoothly with beam energy. In the vicinity of a critical point, however, 
their behavior is governed not simply by non-monotonicity, but by universal 
scaling associated with the growth of a correlation length, modified by the 
finite size and lifetime of the medium.

Identifying the QCD CEP in relativistic heavy-ion collisions is intrinsically 
challenging because these finite, rapidly evolving systems cannot sustain the 
divergent length scales characteristic of equilibrium critical phenomena. 
Instead, the correlation length saturates at a value limited by the system size 
and available evolution time, replacing true phase transitions with smooth 
crossovers and attenuating—or even obscuring—the non-monotonic signatures often 
used as empirical indicators of critical behavior.

Accordingly, the absence of pronounced non-monotonic trends in cumulant ratios 
as a function of beam energy does not imply the absence of critical behavior. 
Finite-size and finite-time effects can suppress or distort critical signatures 
in a correlated manner, rendering background-subtraction approaches unreliable 
in finite, dynamical systems. Because both the putative critical signal and 
non-critical contributions are modified by the same constraints, clean 
separation via model-dependent baselines is not well defined. These limitations 
motivate the use of scaling-based analyses that do not rely on absolute 
magnitudes or subtraction schemes. When finite-time effects are sub-dominant, 
finite-size scaling provides a controlled means of uncovering universal 
behavior; when dynamical limitations dominate the evolution of critical 
fluctuations, extensions based on finite-time scaling may be required
~\cite{Mukherjee:2017kxv}.

A central objective of finite-size scaling (FSS) analyses is therefore to
simultaneously constrain both the location and the universality class of the
QCD critical end point. These two elements are inseparable: a critical point is
physically meaningful only when its coordinates in the QCD phase diagram,
$(T_{\rm CEP}, \mu_{B,\rm CEP})$, specified by the temperature $T$ and baryon
chemical potential $\mu_B$, are identified together with the associated
universality class encoded in the critical exponents. The critical exponents
govern how observables scale in the vicinity of the CEP and provide a
distinctive fingerprint of the underlying critical dynamics. Absent this dual
constraint, an inferred CEP location lacks theoretical significance, while
claims of critical scaling without localization remain experimentally
ambiguous. The FSS framework addresses this requirement by demanding that data
from different system sizes and beam energies collapse onto a universal scaling
function when expressed in appropriate reduced variables, thereby providing a
robust and internally consistent criterion for CEP identification.

Lower-order cumulant ratios such as $C_2/C_1$ and $C_3/C_2$ are particularly
well suited for this purpose. Because they benefit from partial cancellation of
volume-dependent effects and are dominated by fluctuations evolving on shorter
timescales, these ratios are comparatively robust against finite-time
suppression and less sensitive to critical slowing down than higher-order,
non-Gaussian cumulants. As a result, they serve as experimentally accessible
proxies for susceptibility-related quantities. Finite-size scaling studies of
proxy compressibility observables have demonstrated robust scaling behavior
consistent with expectations near the CEP~\cite{Lacey:2014wqa}. By extension,
cumulant ratios such as $C_2/C_1$ and $C_3/C_2$ are expected to exhibit analogous
scaling trends, reinforcing their utility for constraining critical behavior
within the FSS framework.

Non-perturbative QCD configurations, such as baryon junctions, are expected to 
play an important role in shaping net-baryon fluctuations at lower beam 
energies, where baryon stopping and large values of $\mu_B$ become 
significant~\cite{Kharzeev:1996sq,Kharzeev:2000ph,Pihan:2024lxw,Lacey:2024bcm}. 
By facilitating the transport of baryon number to mid-rapidity, baryon 
junctions enhance local baryon density fluctuations and can amplify cumulant 
ratios such as $C_2/C_1$, $C_3/C_2$, and higher-order measures including 
$C_4/C_2$. 
In the vicinity of the CEP, these enhanced density fluctuations may couple to 
critical dynamics, contributing to the growth of compressibility and 
higher-order susceptibilities encoded in the cumulant ratios~\cite{Stephanov:2008qz,
Lacey:2014wqa}.

At the same time, baryon junctions introduce additional non-critical sources of 
fluctuation that can mimic or obscure critical behavior if interpreted solely 
through the magnitude or non-monotonicity of individual observables. This dual 
role underscores the necessity of scaling-based analyses: within a finite-size 
scaling framework, genuinely critical contributions are constrained by 
universality and system-size dependence, whereas junction-driven effects are 
not required to exhibit universal scaling behavior. The ability to distinguish 
between these behaviors provides a key motivation for the scaling approach 
adopted in this work.

Within this context, the power of the finite-size scaling (FSS) approach lies in
the simultaneous and correlated constraints it imposes on both the location and
the universality class of the CEP. When expressed in appropriate reduced
variables, cumulant ratios measured across different beam energies,
centralities, and system sizes are required to collapse onto common scaling
functions. This collapse condition sharply restricts the admissible values of
the critical exponents and the CEP coordinates, providing a stringent test of 
critical consistency that cannot be satisfied by generic non-critical mechanisms.

In this way, FSS does not merely accommodate finite-size and finite-time 
effects, but converts them into quantitative handles that render any 
inferred CEP in the QCD phase diagram both experimentally
credible and theoretically meaningful.

Near the CEP, cumulant ratios are expected to obey FSS relations governed by the 
universality class of the phase transition, which for QCD is commonly associated 
with the three-dimensional Ising (3D Ising) universality class~\cite{Cardy:1996,Stephanov:2008qz}. 
Finite-size scaling analyses of susceptibility-related proxy observables have 
provided empirical support for 3D Ising-like critical behavior in this context~\cite{Lacey:2014wqa}. 
Motivated by these results, the present work extends the FSS framework directly 
to net-baryon cumulant ratios.
 The expected scaling relations take the form
\begin{align}
\frac{C_2}{C_1} &= L^{\gamma/\nu} f_{21}(t L^{1/\nu}, h L^{\Delta/\nu}), \quad
\frac{C_3}{C_2} = L^{-\gamma/\nu} f_{32}(t L^{1/\nu}, h L^{\Delta/\nu}), \nonumber \\
\frac{C_4}{C_2} &= L^{-\gamma/\nu} f_{42}(t L^{1/\nu}, h L^{\Delta/\nu}), \quad
\frac{C_3}{C_1} = L^{-(d-\gamma)/\nu} f_{31}(t L^{1/\nu}, h L^{\Delta/\nu}), \nonumber \\
\frac{C_4}{C_1} &= L^{(d+\alpha)/\nu} f_{41}(t L^{1/\nu}, h L^{\Delta/\nu}),
\label{eq:sf}
\end{align}
where $L$ denotes a characteristic transverse size of the system, and $t$ and $h$ are the reduced temperature and field variables, respectively, that quantify proximity to the CEP.
 Here $d$ denotes the spatial dimensionality, and 
$\gamma$, $\nu$, $\alpha$, and $\Delta$ are the critical exponents associated 
with the susceptibility, correlation length, specific heat, and field-like 
scaling variable, respectively. The functions $f_{ij}$ are universal up to 
normalization.

\begin{figure}
  \includegraphics[width=0.275\textwidth]{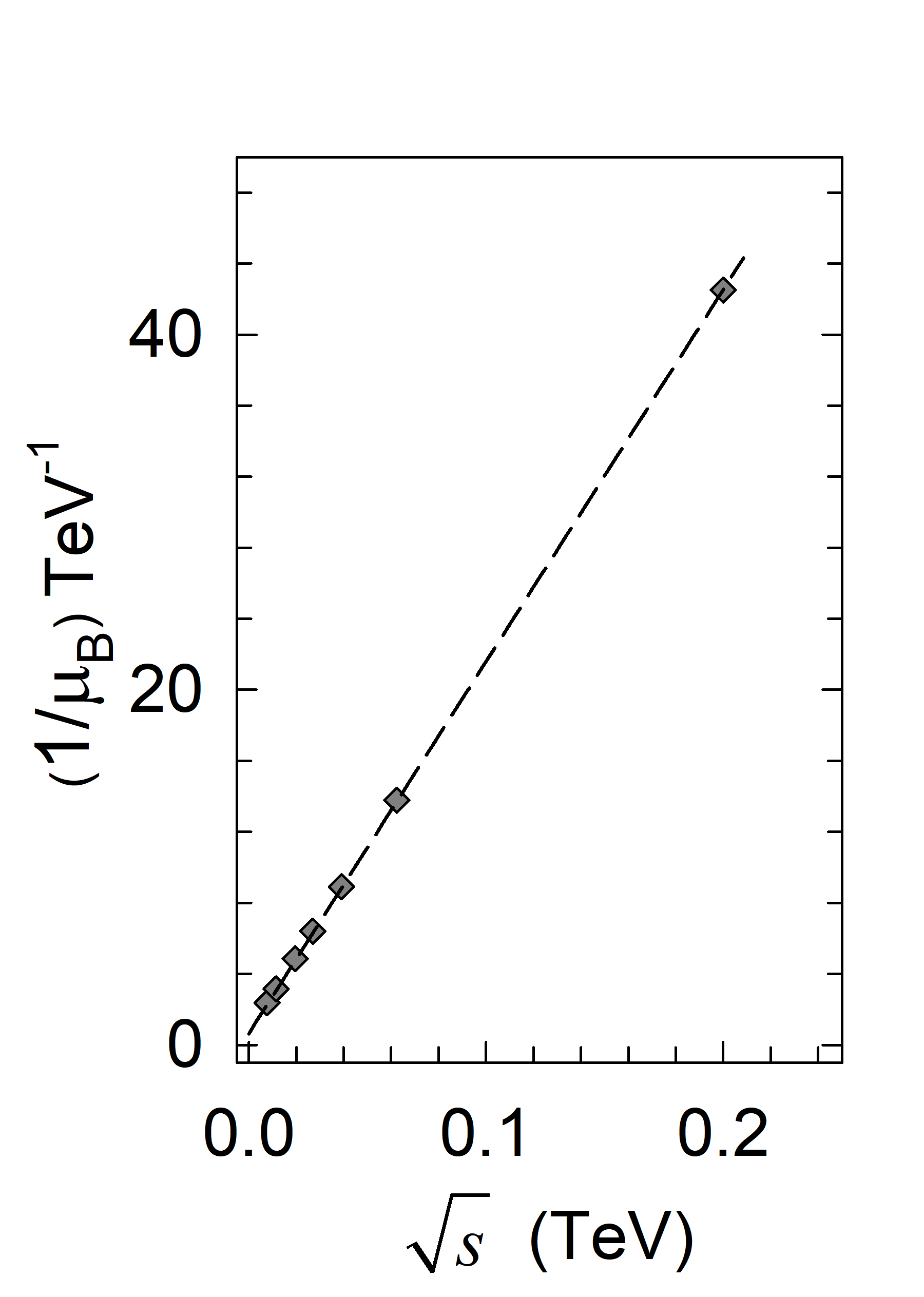}
	\vskip -0.3cm
	\caption{(Color online) Relationship between $1/\mu_B$ and the beam energy 
$\sqrt{s}$. The values of $\mu_B$ are obtained from empirical chemical 
freeze-out parametrizations~\cite{Cleymans:2005xv, Andronic:2009gj}. Over the 
Beam Energy Scan range, $1/\mu_B$ exhibits an approximately linear dependence 
on $\sqrt{s}$, motivating its use as a practical proxy for field-driven scaling 
analyses.}
    \label{fig1}
\end{figure}

Within this framework, applying finite-size scaling across multiple cumulant 
ratios enables nontrivial consistency checks between independent observables and 
strengthens the robustness of the extracted critical parameters. Comparing 
density-driven and field-driven scaling paths further exploits the empirical 
relationship between $\sqrt{s}$ and $1/\mu_B$ (see Fig.~\ref{fig1}) to probe 
thermodynamic conditions near the CEP along complementary directions in the QCD 
phase diagram. The requirement that both scaling trajectories converge on a 
common set of critical parameters imposes a stringent internal consistency 
condition and reinforces a unified interpretation of the observed fluctuation 
behavior.

Cumulant ratios such as $C_2/C_1$, $C_3/C_2$, $C_3/C_1$, $C_4/C_1$, and $C_4/C_2$ 
offer complementary sensitivity to critical dynamics in the vicinity of the 
CEP. Ratios with full volume cancellation, notably $C_2/C_1$ and $C_3/C_2$, are 
particularly well suited for finite-size scaling analyses, as they provide 
intensive probes of susceptibility-driven singular behavior. Of these, 
$C_2/C_1$, which is directly related to the second-order baryon number 
susceptibility, is expected to diverge upward, reflecting enhanced 
compressibility and long-range correlations, while $C_3/C_2$ is predicted to 
diverge downward, signaling the emergence of asymmetry and non-Gaussian 
fluctuations.

Higher-order ratios provide additional, though increasingly constrained, 
information. The volume-cancelled ratio $C_4/C_2$ probes fourth-order 
fluctuations and is predicted to diverge downward, but its scaling fidelity is 
more susceptible to finite-time suppression owing to the slower development of 
higher-order correlations. Ratios with partial volume cancellation, such as 
$C_3/C_1$ and $C_4/C_1$, remain sensitive to the evolving shape of the 
net-baryon distribution, with downward and upward divergences, respectively, 
reflecting changes in skewness and kurtosis.

Taken together, the distinct scaling tendencies of these cumulant ratios enable 
a differential probe of critical behavior. Their combined analysis strengthens 
the finite-size scaling framework by permitting cross-consistency checks across 
observables that emphasize different moments of the underlying fluctuation 
distribution, thereby enhancing the robustness of the extracted CEP 
constraints.

The system-size parameter $L\equiv\bar{R}$ used in the finite-size scaling 
analysis is obtained from Monte Carlo Glauber (MC-Glauber) simulations 
\cite{Miller:2007ri,Lacey:2010hw} performed over a range of collision 
centralities and beam energies. In this framework, nucleons undergoing at least 
one inelastic nucleon--nucleon interaction define the participant set 
$N_{\rm part}$. The transverse spatial distribution of these participants in 
the $x$--$y$ plane is characterized by root-mean-square widths $\sigma_x$ and 
$\sigma_y$ along the principal axes of the overlap region, from which the 
characteristic transverse size is defined as 
$1/\bar{R}=\sqrt{(1/\sigma_x^2)+(1/\sigma_y^2)}$, following 
Ref.~\cite{Bhalerao:2005mm}.

For finite-size scaling, the essential input is the \emph{relative variation} 
of $L$ across beam energies and collision centralities, rather than its 
absolute magnitude. As a result, the scaling analysis is largely insensitive to 
overall normalization uncertainties in $\bar{R}$. The physical relevance of 
$\bar{R}$ as a system-size proxy is further supported by its strong empirical 
correlation with interferometric HBT radii ($R_{\rm out}$, $R_{\rm side}$, and 
$R_{\rm long}$), which characterize the space--time extent of the particle-
emitting source at kinetic freeze-out~\cite{Lacey:2014rxa,Adare:2014qvs}. The 
approximately linear dependence of these radii on $\bar{R}$ over a broad range 
of beam energies indicates that the initial transverse geometry reliably 
tracks the final freeze-out volume.

Systematic uncertainties in $\bar{R}$, dominated by variations in Glauber model 
inputs, are estimated to be at the few-percent level~\cite{Lacey:2010hw}. Taken 
together, these considerations establish $\bar{R}$ as a robust and physically 
well-motivated measure of system size for finite-size scaling analyses of 
critical phenomena.

\begin{figure*}
  \includegraphics[width=0.75\textwidth]{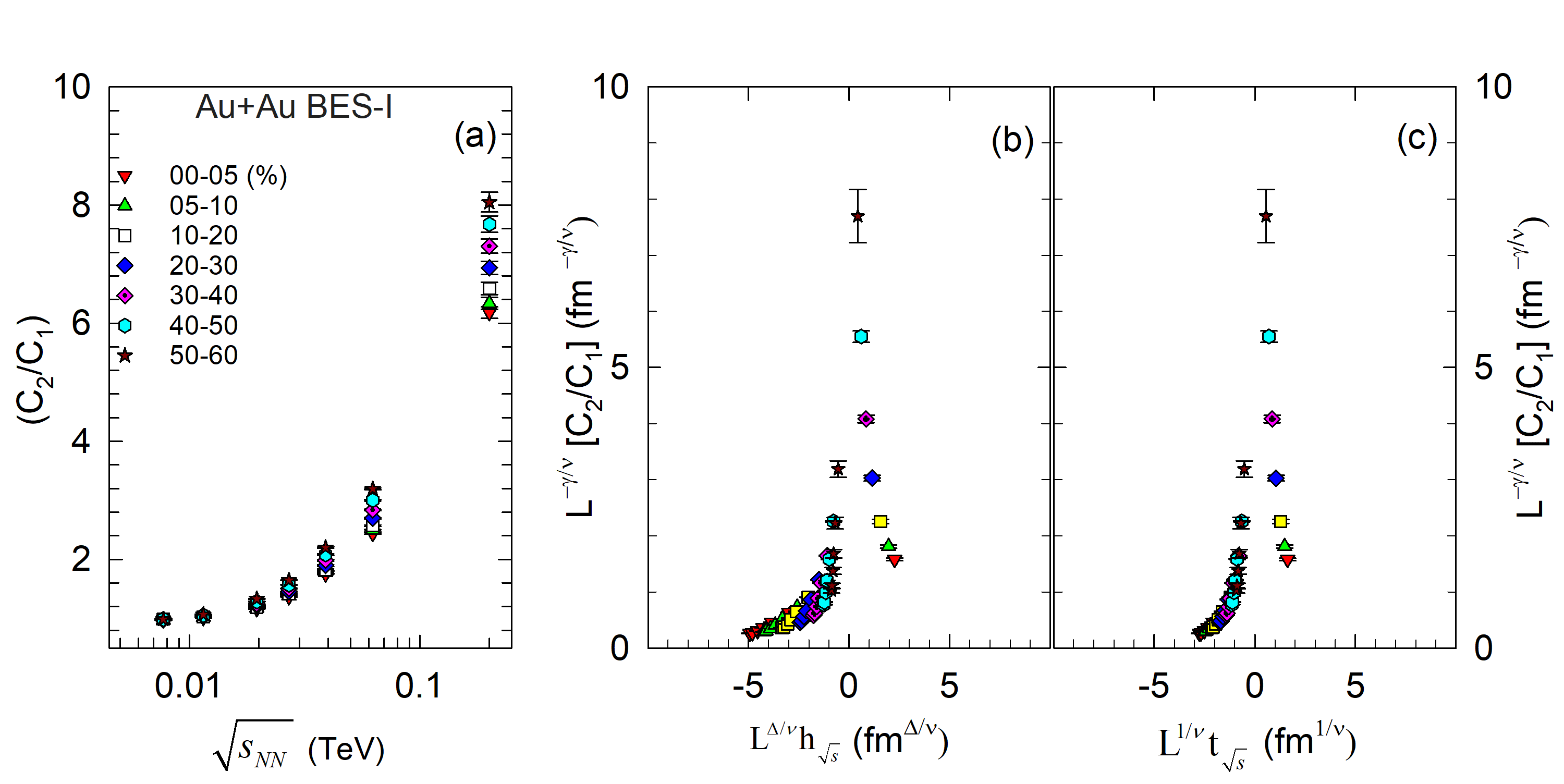}
	\vskip -0.3cm
  \caption{(Color online)
Illustration of the finite-size scaling procedure.
Panel (a) shows the beam-energy dependence of the unscaled cumulant ratio
$C_2(\mathrm{cent})/C_1(\mathrm{cent})$ for Au+Au collisions across the indicated
centrality intervals. Panels (b) and (c) show the corresponding field-driven and
density-driven finite-size scaling functions. The collapse of data from
different system sizes in the scaled representations reveals the upward
divergence expected for $C_2/C_1$, consistent with a compressibility-related
observable in the vicinity of the CEP.
}
\label{fig2}
\end{figure*}
\begin{figure*}
  \includegraphics[width=0.75\textwidth]{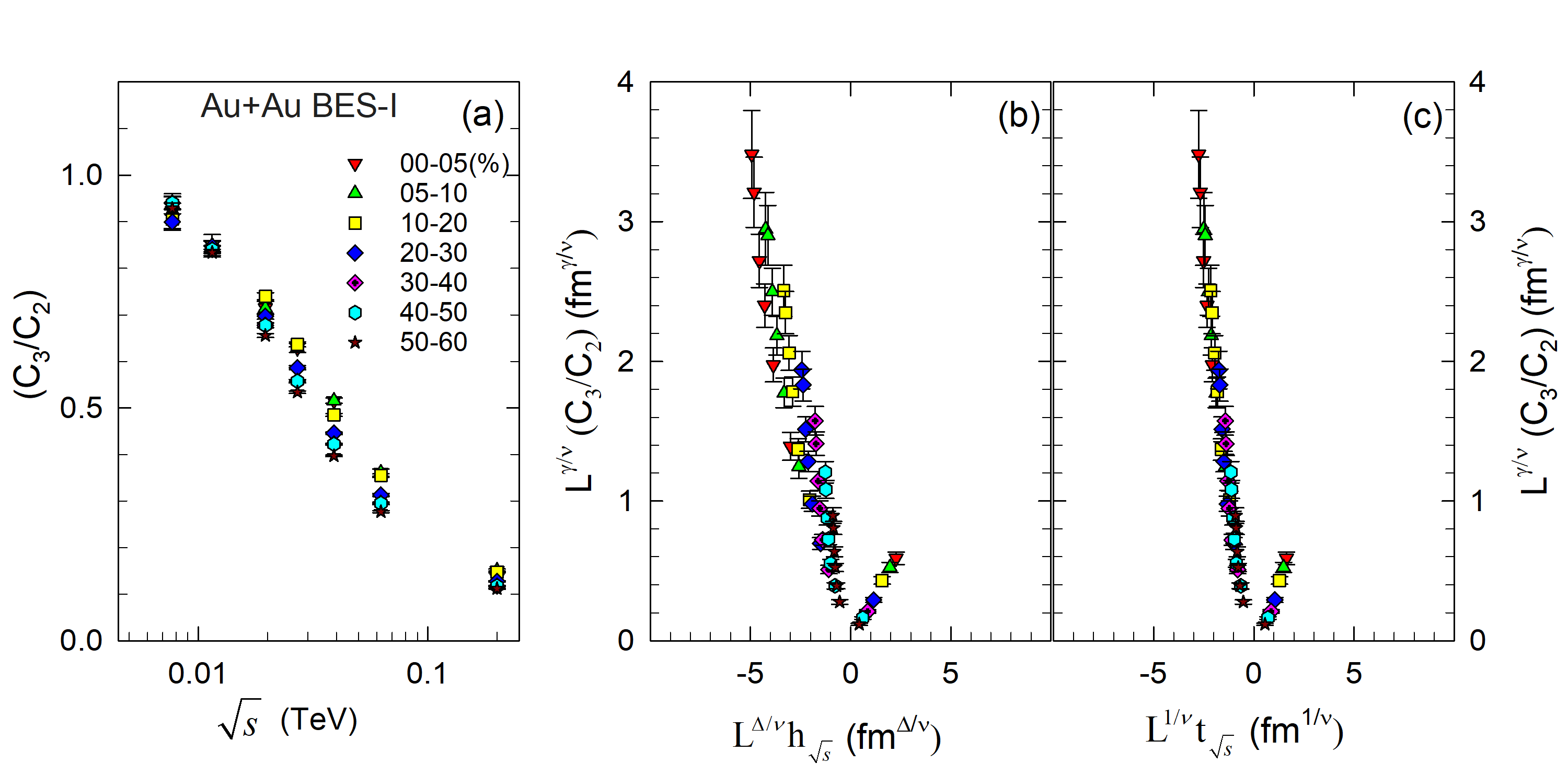}
	\vskip -0.3cm
	\caption{(Color online)
Same as Fig.~2, but for the cumulant ratio $C_3(\mathrm{cent})/C_2(\mathrm{cent})$. 
Panel (a) shows the unscaled beam-energy dependence across centralities, while 
panels (b) and (c) display the corresponding field-driven and density-driven 
scaling functions. The scaled data exhibit the downward divergence expected for 
$C_3/C_2$, consistent with the emergence of skewness near the CEP.
}
    \label{fig3}
\end{figure*}
\begin{figure*}
  \includegraphics[width=0.75\textwidth]{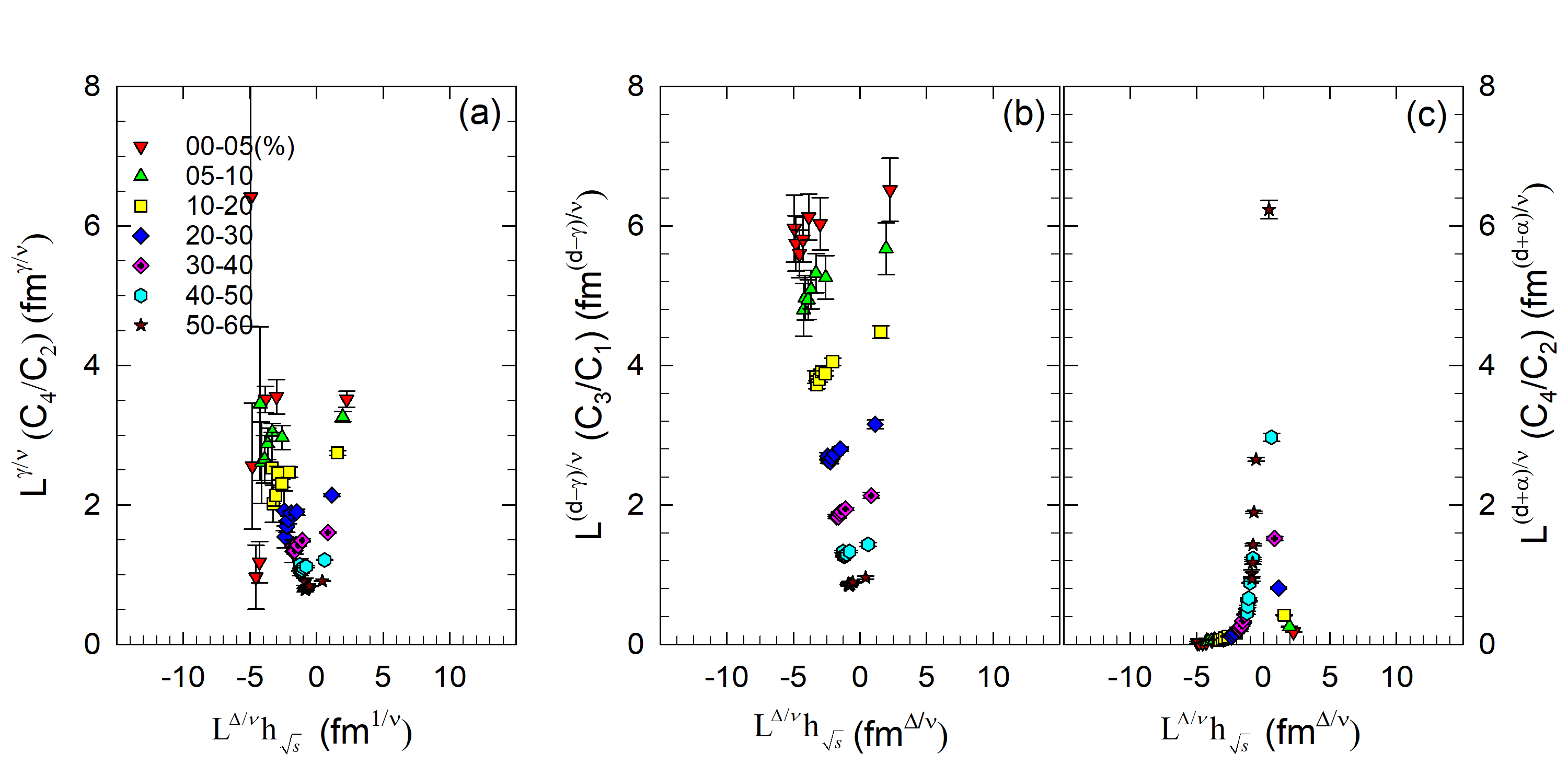}
	\vskip -0.3cm
	\caption{(Color online)
Field-driven finite-size scaling functions for the cumulant ratios 
$C_4(\mathrm{cent})/C_2(\mathrm{cent})$ (a), $C_3(\mathrm{cent})/C_1(\mathrm{cent})$ (b), 
and $C_4(\mathrm{cent})/C_1(\mathrm{cent})$ (c). The scaled data exhibit distinct 
divergence patterns—downward for $C_4/C_2$ and $C_3/C_1$, and upward for 
$C_4/C_1$—consistent with expectations for 3D Ising-like critical behavior in 
the vicinity of the CEP.
}
    \label{fig4}
\end{figure*}

To implement finite-size scaling (FSS), the cumulant ratios
$C_2/C_1$, $C_3/C_2$, $C_4/C_2$, $C_3/C_1$, and $C_4/C_1$ measured in Au+Au
collisions over the full Beam Energy Scan Phase~I range
($\sqrt{s}=7.7$--$200$~GeV)~\cite{STAR:2021iop} are analyzed across multiple
centralities. This data set spans a broad range of system sizes and beam
energies, providing the leverage required to test scaling behavior in the
vicinity of the CEP.

The analysis employs critical exponents from the 3D Ising universality class
($\nu=0.630$, $\gamma=1.237$, $\beta=0.326$, and $\Delta=1.563$), which govern the
system-size and control-parameter scaling of the cumulant ratios
(cf.~Eq.~\ref{eq:sf}). A successful FSS description is signaled by the collapse
of rescaled data for all cumulant ratios, measured across different beam
energies and centralities, onto common scaling curves, thereby imposing
simultaneous constraints on both the CEP location and its universality class.

Because the reduced temperature $t$ and external field $h$ are not directly
accessible experimentally, the beam energy $\sqrt{s}$ is used as a practical
proxy. Parametrizations of the chemical freeze-out curve
\cite{Cleymans:2005xv,Andronic:2009gj} enable a mapping to thermodynamic
coordinates $(T,\mu_B)$ and the construction of two complementary scaling
variables corresponding to field-driven and density-driven trajectories in the
QCD phase diagram. The field-driven trajectory emphasizes net-baryon
fluctuations, which are particularly sensitive to $\mu_B$ and are therefore
expected to exhibit enhanced sensitivity to critical scaling.

\noindent{\em Field-driven scaling variable $h_{\sqrt{s}}$:}
Net-baryon fluctuations, dominated by field-like dynamics and highly sensitive
to variations in $\mu_B$~\cite{Asakawa:2000wh,Stephanov:2011zz}, are analyzed
using the field-driven scaling variable
\[
h_{\sqrt{s}}=
\frac{(1/\mu_B)-(1/\mu_{B,\rm CEP})}{(1/\mu_{B,\rm CEP})}.
\]
Over the beam-energy range of interest, $1/\mu_B$ exhibits an approximately
linear dependence on $\sqrt{s}$, as illustrated in Fig.~\ref{fig1} based on
standard freeze-out parametrizations~\cite{Cleymans:2005xv,Andronic:2009gj}.
Finite-size scaling with $h_{\sqrt{s}}$ enables a direct extraction of the
critical beam energy $\sqrt{s}_{\rm CEP}$, which fixes the CEP location along
the experimentally accessible trajectory. This value is subsequently mapped to
$\mu_{B,\rm CEP}$ and $T_{\rm CEP}$ using the same freeze-out parametrization.

While the scaling collapse obtained with $h_{\sqrt{s}}$ is robust for
determining $\sqrt{s}_{\rm CEP}$, reinserting the inferred $\mu_{B,\rm CEP}$
into the scaling variable introduces additional uncertainty, primarily due to
systematic effects in the freeze-out mapping. Consequently, $\sqrt{s}$ serves
as the more stable experimental control variable for field-driven scaling, with
$\mu_{B,\rm CEP}$ retained for thermodynamic interpretation.

\vspace{0.5em}
\noindent{\em Density-driven scaling variable $t_{\sqrt{s}}$:}
Density-driven scaling is implemented using the reduced beam-energy variable
\[
t_{\sqrt{s}}=\frac{\sqrt{s}-\sqrt{s}_{\rm CEP}}{\sqrt{s}_{\rm CEP}},
\]
which treats $\sqrt{s}_{\rm CEP}$ as the experimentally accessible critical
parameter. Because $\mu_B$ varies approximately inversely with $\sqrt{s}$ over
the relevant energy range (Fig.~\ref{fig1}), variations in $\sqrt{s}$ primarily
reflect changes in baryon density, while the temperature remains nearly
constant. The value of $\sqrt{s}_{\rm CEP}$ determined from the field-driven
scaling analysis is used consistently in the density-driven construction,
thereby enforcing internal consistency between the two scaling trajectories and
allowing $t_{\sqrt{s}}$ to quantify deviations from criticality along a
density-driven path.

Together, $h_{\sqrt{s}}$ and $t_{\sqrt{s}}$ generate complementary scaling
functions along distinct trajectories in the QCD phase diagram. Both scaling
analyses are governed by the same set of critical exponents fixed by the
3D Ising universality class, with internal consistency enforced by the
requirement that both collapses occur for a common CEP location.

Figures~\ref{fig2} and \ref{fig3} illustrate the finite-size scaling behavior of
$C_2/C_1$ and $C_3/C_2$, respectively. Panels~(a) show that finite-size and
finite-time effects obscure the non-monotonic features often expected near a
critical point when the data are examined in their raw form. In contrast,
panels~(b) and (c) demonstrate that, once scaled within the finite-size scaling
(FSS) framework, measurements from different beam energies and centralities
collapse onto a common scaling curve, revealing the underlying critical
behavior. The resulting divergence patterns are consistent with expectations
for 3D Ising-like critical behavior: $C_2/C_1$, related to compressibility,
exhibits an upward divergence, while $C_3/C_2$, associated with skewness,
diverges downward.

A modest difference in scaling fidelity is observed between the field-driven and
density-driven representations, with slightly better collapse obtained for the
density-driven case, consistent with the additional sensitivity of the
field-driven variable to uncertainties in the $\sqrt{s}\!\leftrightarrow\!\mu_B$
mapping.

A key implication of these results is that finite-size scaling does not require
visible non-monotonicity of the \emph{unscaled} or background-subtracted
cumulant ratios as a function of beam energy within individual centrality bins.
Apparent monotonic behavior of the raw data is therefore not in conflict with
critical dynamics in finite, dynamically evolving systems. Instead, the
defining signature of proximity to the QCD critical end point (CEP) is the
collapse of data from systems of different sizes and beam energies onto a
universal scaling function when expressed in appropriate reduced variables,
together with scaling functions that exhibit the divergence patterns expected
for the underlying universality class. Since non-monotonic trends in unscaled or
background-subtracted observables can arise from non-critical effects, they do
not constitute a reliable diagnostic of criticality in finite dynamical
systems. Rather, the emergence of universal scaling across system sizes—consistent
with the expected critical divergences—provides a robust and model-independent
indicator of critical behavior.

Figure~\ref{fig4} shows the field-driven scaling functions for $C_4/C_2$,
$C_3/C_1$, and $C_4/C_1$. The consistent collapse observed across these cumulant
ratios—each probing different moments of the net-baryon distribution—indicates
that the scaling behavior reflects proximity to a common critical region rather
than statistical or procedural artifacts. Each ratio exhibits a distinct
divergence pattern consistent with expectations for 3D Ising-like critical
behavior: $C_4/C_2$, sensitive to non-Gaussian kurtosis, diverges downward;
$C_3/C_1$, reflecting skewness relative to the mean, also diverges downward;
while $C_4/C_1$, which combines kurtosis- and compressibility-driven
contributions, diverges upward. The internal consistency of these complementary
trends provides a nontrivial cross-check of the scaling framework and supports a
unified critical interpretation.

All scaling collapses shown in Figs.~\ref{fig2}–\ref{fig4} are achieved using a
single, fixed set of critical exponents from the 3D Ising universality class,
applied uniformly across all observables (cf.~Eq.~(\ref{eq:sf})). Importantly,
although a single universality class is assumed, different cumulant ratios
depend on these exponents through distinct scaling powers. The ability to
achieve simultaneous scaling collapse across multiple independent ratios using
one CEP location and one exponent set therefore constitutes a stringent
internal consistency test of 3D Ising-like critical behavior; such
overconstrained agreement would not, in general, be expected for an incorrect
universality assignment.

The extracted CEP coordinates are therefore constrained primarily by the
requirement of correlated scaling collapse across multiple observables and
scaling trajectories, rather than by a single fit or extremum. The stability of
the collapse against variations in the system-size parameter $L$, the choice of
scaling variables, and commonly used freeze-out parametrizations demonstrates
that the inferred CEP location is robust within these systematic uncertainties.
In particular, variations in the freeze-out mapping lead to shifts of order
$\pm5$~MeV in $T_{\rm CEP}$ and $\pm10$~MeV in $\mu_{B,\rm CEP}$, without
degrading the overall scaling consistency.

The modest variation in scaling fidelity among different cumulant ratios is
consistent with the expected influence of finite-time effects, which can
differentially suppress higher-order fluctuations without eliminating the
finite-size scaling behavior. The persistence of a common scaling collapse
across observables indicates that finite-size effects remain dominant, placing
the system in a regime where finite-size scaling is applicable. Taken together,
these observations demonstrate the internal consistency of the analysis and
support an interpretation of the observed fluctuations in terms of 3D
Ising-like critical behavior in the vicinity of the QCD critical end point.

At lower beam energies, where baryon stopping is substantial and the baryon
chemical potential is large, non-equilibrium baryon-number transport mechanisms
are expected to play an increasingly important 
role~\cite{Kharzeev:1996sq,Kharzeev:2000ph,Pihan:2024lxw,Lacey:2024bcm}. 
In particular, non-perturbative QCD configurations such as baryon junctions can 
generate long-wavelength, event-by-event baryon-density fluctuations that populate the
same susceptibility channels probed by the cumulant ratios analyzed here. Near
the CEP, such dynamically generated fluctuations may couple to incipient
critical modes, effectively enhancing the experimental visibility of critical
scaling in a finite, rapidly evolving medium. This mechanism does not modify
the underlying universality class, but can amplify susceptibility-driven
signals in ways that are not captured by equilibrium approaches. In this sense,
dynamical baryon transport provides a natural bridge between finite-system
experimental observations and equilibrium lattice calculations, which may show
muted or inconclusive CEP signatures at comparable values of $\mu_B$.

In summary, a finite-size scaling analysis of the cumulant ratios
$C_2/C_1$, $C_3/C_2$, $C_4/C_2$, $C_3/C_1$, and $C_4/C_1$ measured in Au+Au
collisions over the full BES-I energy range reveals robust scaling behavior
consistent with proximity to the QCD critical end point (CEP). When analyzed
using both field-driven and density-driven scaling variables, data from
different system sizes and beam energies collapse onto common scaling functions,
exhibiting distinct divergence patterns characteristic of 3D Ising-like
critical behavior: upward for $C_2/C_1$ and $C_4/C_1$, and downward for
$C_3/C_2$, $C_3/C_1$, and $C_4/C_2$. The extracted CEP coordinates,
$\sqrt{s}_{\rm CEP}\approx33.0$~GeV, $\mu_{B,\rm CEP}\approx130$~MeV, and
$T_{\rm CEP}\approx158.5$~MeV, are stable against reasonable variations in
freeze-out parametrizations and system-size estimates and are constrained
within a single, self-consistent universality class. In this finite, dynamically
evolving system, the defining signature of criticality is the emergence of
universal scaling across observables—together with the expected divergence
structure of the scaling functions—rather than non-monotonic behavior in raw or
background-subtracted data. The observed hierarchy in scaling fidelity indicates
that finite-size effects dominate, while finite-time effects modulate but do not
eliminate the scaling behavior. Dynamical baryon-number transport mechanisms,
such as baryon junctions, can seed long-wavelength baryon-density fluctuations
that couple to incipient critical modes, enhancing the experimental visibility
of critical scaling without altering the underlying universality class. This
provides a natural framework for understanding why critical behavior may be
experimentally accessible in heavy-ion collisions even when equilibrium lattice
QCD calculations, limited by finite volume, analytic continuation, and the
absence of real-time baryon transport, yield weak or inconclusive constraints
on the CEP at comparable values of $\mu_B$.

%
\bibliography{FS_Scaling-refs-BES}
%
\end{document}